# ST-NeRP: Spatial-Temporal Neural Representation Learning with Prior Embedding for Patient-specific Imaging Study

Liang Qiu, Liyue Shen, Lianli Liu, Junyan Liu, Yizheng Chen, and Lei Xing

*Abstract*—During and after a course of therapy, imaging is routinely used to monitor the disease progression and assess the treatment responses. Despite of its significance, reliably capturing and predicting the spatial-temporal anatomic changes from a sequence of patient-specific image series presents a considerable challenge. Thus, the development of a computational framework becomes highly desirable for a multitude of practical applications. In this context, we propose a strategy of Spatial-Temporal Neural Representation learning with Prior embedding (ST-NeRP) for patient-specific imaging study. Our strategy involves leveraging an Implicit Neural Representation (INR) network to encode the image at the reference time point into a prior embedding. Subsequently, a spatial-temporally continuous deformation function is learned through another INR network. This network is trained using the whole patient-specific image sequence, enabling the prediction of deformation fields at various target time points. The efficacy of the ST-NeRP model is demonstrated through its application to diverse sequential image series, including 4D CT and longitudinal CT datasets within thoracic and abdominal imaging. The proposed ST-NeRP model exhibits substantial potential in enabling the monitoring of anatomical changes within a patient throughout the therapeutic journey.

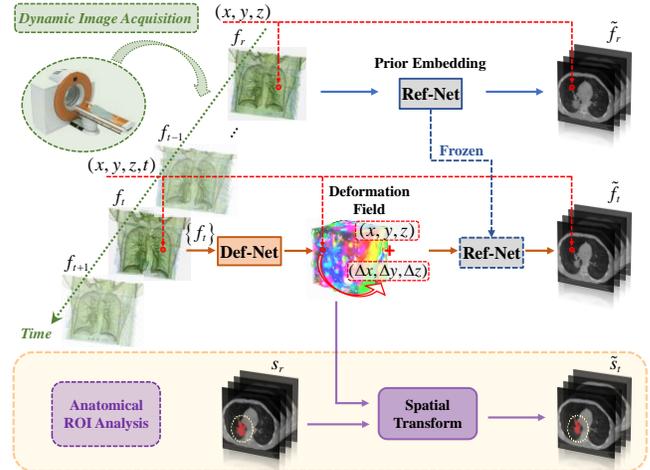

**Fig. 1.** Overview of ST-NeRP for patient-specific study.

## I. INTRODUCTION

Longitudinal imaging studies, which involve repeated examinations of the same individual, are pivotal for understanding disease progression and optimizing patient management strategies over the course of treatment. By monitoring anatomical changes across sequential imaging scans, clinicians gain invaluable insights into how diseases evolve and how treatments impact them. This allows for timely interventions and personalized adjustments to treatment plans. However, manual analysis of these imaging studies is labor-intensive, time-consuming, and prone to variability between observers due to data complexity and patient diversity. Therefore, there is a growing demand for computer-aided tools to streamline the analysis process, reduce costs, accelerate turnaround times, and enhance overall reliability. Such tools have the potential to greatly improve patient care by providing more accurate and timely assessments of disease progression and treatment response.

Image registration serves as a critical approach in longitudinal analysis, enabling the tracking of changes or evolution in anatomical regions like tumors or lesions over time. Traditional optimization-based methods often rely on predefined mathematical or computational models to describe image transformations, such as rigid or elastic deformations. While effective in many cases, these approaches can be time-consuming and suffer from limited accuracy, particularly when dealing with less distinctive image features [1]. Recent advances in learning-based registration algorithms offer promising solutions by directly learning complex patterns and relationships from data, potentially overcoming these limitations across a wide range of applications and imaging modalities [2]. Additionally, automatic image-based anatomical monitoring can be approached through (3D) object segmentation, detection and tracking, leveraging sophisticated feature representation techniques, an area of interest for both computer vision and medical image analysis communities [3-9]. A recent study introduced a deep learning algorithm integrating appearance-based feature representation and anatomical constrains to track lesion motion within 4D longitudinal images, primarily focusing on monitoring changes in the center and radius of lesions [10]. To enhance the effectiveness of the representation embeddings, a multi-

This work was supported by the National Institutes of Health (NIH) (1R01CA223667, 1R01CA176553, and 1R01CA227713). (Liang Qiu and Liyue Shen are co-first authors.) (Corresponding author: Lei Xing.)

Liang Qiu, Lianli Liu, Junyan Liu, Yizheng Chen and Lei Xing are with the Department of Radiation Oncology, Stanford University, Stanford, CA 94305 USA. (E-mail: qiuliang@stanford.edu; lliu@stanford.edu; junyan01@stanford.edu; chenyz@stanford.edu; lei@stanford.edu).

Liyue Shen is with the Department of Electrical Engineering and Computer Science, University of Michigan, Ann Arbor, MI 48109 USA. (E-mail: liyues@umich.edu).

scale self-supervised learning method was proposed by learning a similarity map between multi-timepoint image acquisitions using a pixel-wise contrastive learning strategy. This approach enables the generic application to different types of lesions and image modalities [11]. For a more comprehensive analysis of complex lesion changes, such as lesion merging or splitting in the longitudinal study, a graph-theoretic lesion tracking method was developed [12]. Nevertheless, those methods fall short in capturing temporally continuous changes within patient-specific sequences, and necessitate extensive data for model training, leading to significant labor and time demands and potential domain shift in medical imaging domain.

Recently, implicit neural representation (INR) [13-20] learning has emerged as a data-efficient approach for learning patient-specific representations [21], with its unique advantages explored across various medical imaging tasks, including image registration [22, 23], segmentation [24, 25], reconstruction [21, 26], high/super-resolution [27, 28] and shape synthesis [29]. In this paper, we investigate the potential of the technique in learning the spatiotemporal representation for patient-specific imaging studies (see Fig. 1). Notably, our model, unlike previous deformable registration methods that typically establish correspondence solely at two distinct time points, can predict unseen images at any target time point and estimate anatomical structures simultaneously. The essence of the approach is leveraging INR to capture the complex spatiotemporal deformation function within patient-specific image sequences.

Our main contributions are summarized as follows:

1) We provide a novel solution for patient-specific imaging study via INR learning. To achieve this, an ST-NeRP model is designed to learn a spatial-temporal deformation function to represent the content of a sequential image series via prior embedding learned from reference image.

2) We present extensive experiments of continuous image interpolation and deformable registration for patient-specific temporal image series, including longitudinal and 4D CT datasets in thoracic and abdominal regions, which substantiate the effectiveness and versatility of our proposed method.

## II. RELATED WORK

### A. INR Learning

INR learning aims to represent an image or high-dimensional data via neural network parametrization. It offers an efficient way to represent 3D geometry and scene [13-20]. For example, DeepSDF utilized signed distance functions (SDFs) to represent 3D shapes, allowing the neural network to approximate the SDF and reconstruct or generate novel 3D shapes [18]. Similarly, NeRF introduced an approach to learn implicit representations of 3D scene, using a neural network to represent the scene's appearance and geometry implicitly, resulting in high-quality view synthesis and realistic renderings [19]. Moreover, INR has gained significant attention in the context of time-series data [30-33]. By incorporating time as an additional input, NeRF was extended to a dynamic domain, enabling the reconstruction and rendering of novel images of objects undergoing rigid and non-rigid motions, all from a single moving camera view [30]. Additionally, a novel method called Occupancy Flow was proposed for the spatiotemporal representation of time-changing 3D geometry, which effectively learns a continuous vector field to estimate the motion temporally and spatially for dense 4D reconstruction [31]. More recently, VideoINR was introduced as a continuous video representation, allowing the sampling and interpolation of the video frames at arbitrary frame rate and spatial resolution simultaneously [32]. Moreover, a framework was proposed which jointly learns INRs from RGB frames and events. This framework enables arbitrary scale video super-resolution by utilizing high temporal resolution property of events to complement RGB frames through spatial-temporal fusion and temporal filter. It also utilizes an effective spatial-temporal implicit representation to recover frames at arbitrary scales [33].

Despite the considerable efforts on neural representation pertaining to natural or photographic images, limited research has been conducted in the medical domain [22, 25, 29, 34]. In [25], an Implicit Organ Segmentation Network (IOSNet) was proposed, which involved learning continuous segmentation functions, diverging from prevailing approaches that rely on discrete pixel/voxel-based representations. IOSNet demonstrated exceptional memory efficiency and accurately delineated organs regardless of their sizes. Another innovative approach utilized a curvature-enhanced implicit function network to generate high-quality tooth models from CBCT images. This method, detailed in [34], employed a curvature enhancement module as a guide for refining the surface reconstruction process, resulting in outstanding tooth model quality. Certain studies have explored medical applications in the spatiotemporal domain, such as cell tracking and segmentation in biomedical images. In this context, an INR model was designed to enable the synthesis of authentic cell shapes within the spatiotemporal domain, conditioning its generation on a latent code [29]. Likewise, a spatiotemporal INR network was put forward to construct a personalized model for monitoring the abdominal aortic aneurysms (AAA) progression. This model achieved proper interpolation of AAA shape by representing the AAA surface as the zero-level set of its signed distance function, rather than operating directly within the image domain [35]. Another relevant study by Wolterink et al. [22] utilized a multi-layer perceptron (MLP) network IDIR to implicitly represent a transformation function, facilitating deformable registration of 4D chest CT series. However, this work focused exclusively on the spatial domain, without considering temporal information.

### B. Deformable Registration of Medical Images

Deformable registration plays a significant role in medical image analysis and longitudinal study, aiming at aligning two or more images by estimating a deformation field that establishes a spatial mapping between corresponding anatomical structures. Despite numerous models developed to address the challenges associated with deformable registration [36-39], their practical applications are often hindered by limited model accuracy, intensive computational resources, and slow runtime. Deep learning techniques have shown great promise in improving computational efficiency and accuracy in this domain. For instance, an encoder-decoder network called Quicksilver was proposed to estimate the deformation field of brain MR images [40]. Balakrishnan et al. introduced VoxelMorph, a comprehensive unsupervised deformable registration network designed for both unimodal and multimodal scenarios. The framework enables the estimation of a dense deformation field in a single step by integrating U-Net and Spatial Transformer Layer (STL), demonstrating significant potential across various applications [41]. Additionally, the exploration of weakly-supervised registration [42, 43] and joint learning of registration and segmentation [44, 45] has revealed promising results, demonstrating improved registration through the integration of auxiliary segmentation or keypoint correspondence. Nonetheless, these methods often rely on substantial training datasets to achieve optimal performance, potentially introducing biases and domain shifts. Such challenges are particularly pertinent in the context of individualized therapy, especially when encountering new patients whose data falls outside the training distribution. While transfer learning and domain adaptation techniques offer avenues for bolstering generalization, they may necessitate additional labeled data or fine-tuning efforts [46]. In addressing this issue, a one-shot learning approach was proposed by Fechter et al. for deformable medical image registration. This approach directly estimates the deformation vector field by optimizing a convolution neural network, proving effective for periodic organ motion tracking [47]. A recent breakthrough by Wolterink et al. introduced an implicit deformable image registration model IDIR to estimate the respiratory motion between two phases in a 4D chest CT dataset. This study showcases the practicality of the model in 4D imaging applications, utilizing data from 10 patients [22]. Van Harten et al. further enhanced IDIR by integrating cycle-consistency to regularize the optimization process, enhancing the robustness and accuracy of the predicted deformation field [48]. Furthermore, Han et al. proposed a cascaded image registration framework called Vas-DNVF to merge the advantages of optimization and data-driven methods, where a fully convolutional neural network was trained to predict the initial deformation, followed by an INR based registration algorithm for further refinement. This strategy enhanced generalizability and achieved precise alignment on large-scale 3D MR brain scan datasets [49].

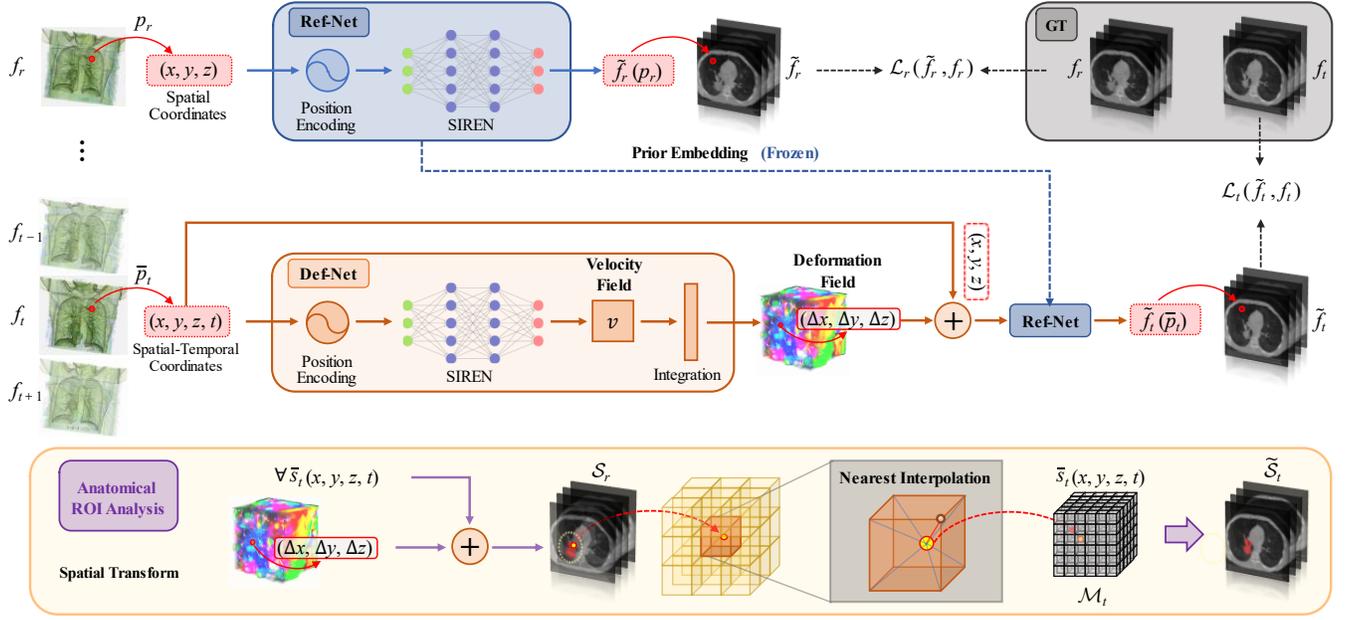

**Fig. 2.** Framework of the proposed ST-NeRP. We first train the coordinate-based reference network Ref-Net to embed the prior image at the reference time point. Then we learn the spatial-temporal deformation network Def-Net to predict the time-dependent deformation field at target time points. Finally, we can perform the anatomical ROI analysis based on the deformation field predicted by Def-Net.

## IV. METHODOLOGY

To learn spatiotemporal deformation of anatomy, we train an ST-NeRP model using a patient's temporal image series. Our method, illustrated in Figure 2, consists of three submodules: (i) reference prior embedding, (ii) spatial-temporal learning, and (iii) patient-specific ROI adaptation. The primary objective is to capture temporal changes in the patient anatomy, particularly in some Regions of Interest (ROIs).

Consider a sequence of medical images, e.g., longitudinal image data, for a specific patient, denoted as $\{f_1, f_2, \cdots, f_T\}$, acquired at time points $\{t_1, t_2, \cdots, t_T\}$. In clinical practice, variations in patient positioning, anatomical changes, or disparities in imaging protocols across diverse scanners and sites can induce substantial misalignments and fluctuations in intensity profiles across longitudinal series, which can significantly impact subsequent image analysis. To mitigate this issue, we assume all the longitudinal images undergo intensity normalization and affine alignment as a preprocessing step, so that the only source of misalignment between the images is nonlinear. First, a reference image $f_r$ obtained from a prior scan of the subject is embedded with an INR, namely prior embedding. This is accomplished by encoding of the entire spatial image domain into the parameters of the reference network (Ref-Net). Ref-Net is trained to learn a continuous function capable of accurately associating spatial coordinates with their corresponding intensity values in the reference image. While it is theoretically feasible to select any image from the patient's longitudinal sequence as the reference, in clinical practice, we typically opt for the initial image. This preference stems from the fact that the first image often contains essential treatment planning information, which can be effectively utilized for subsequent treatment adaptation. Next, the subsequent images collected from the scans $\{f_t\}$ at different time points are exploited to train a deformation network (Def-Net). The introduction of Def-Net is to effectively capture the coordinate displacements and enables the identification of coordinate-intensity correspondences based on the previously established prior embedding, represented by the trained parameters of Ref-Net. The optimization process of Def-Net occurs within the spatial-temporal continuous domain, facilitating the comprehensive acquisition of deformation information. Finally, with the acquired spatial-temporal representation of the patient anatomy, we can adapt the patient image information from the reference time point to any target time point for downstream applications.

*A. Ref-Net: Reference Prior Embedding*

As depicted in Fig. 2, we initiate the learning process of ST-NeRP by representing the reference image $f_r$ as the INR, which is a continuous function parameterized by the Ref-Net $\mathcal{R}_\theta$, defined as

$$\mathcal{R}_\theta : p_r \to f_r(p_r), \text{with } p \in [0,1)^3, f_r(p_r) \in \mathbb{R}, \quad (1)$$

where the input $p_r(x,y,z)$ indicates normalized spatial grid coordinates and the output $f_r(p_r)$ denotes the corresponding intensity value in $f_r$. Fourier mapping, which has demonstrated its efficacy in high-frequency low-dimensional regression tasks [50], is first used to encode the input coordinates $p_r$ before directly fed into an MLP network, shown as follows:

$$\gamma(p) = \left[\cos(2\pi B p), \sin(2\pi B p)\right], \quad (2)$$

where matrix $B$ contains the coefficients for Fourier feature transformation in the position encoding. The entries of matrix $B$ are randomly drawn from Gaussian distribution $\mathcal{N}(0, \sigma^2)$, characterized by a mean of 0 and a standard deviation $\sigma$. Following Fourier feature embedding, the encoded coordinates $\gamma(p)$ are passed into a coordinate based MLP within $\mathcal{R}_\theta$ to parameterize the continuous function representing the reference image. In this study, we utilize SIREN [51] as the MLP backbone for INR, leveraging periodic activation functions that enable the model to capture intricate signal features effectively. Notably, it can also be adapted to other backbone models.

During the network training, the objective function is optimized using the following formulation:

$$\theta^* = \arg\min \frac{1}{N} \Sigma_{i=1}^{N} \| \mathcal{R}_\theta(p_r^i) - f_r(p_r^i) \|_2^2, \quad (3)$$

where the minimization is performed over all the coordinate-intensity pairs $\{p_r^i, f_r(p_r^i)\}_{i=1}^{N}$ within $f_r$, totaling $N$ pixels. After training, the optimized reference network $\mathcal{R}_{\theta^*}$ with the corresponding network parameters $\theta^*$ embeds the intrinsic information of $f_r$ as a prior. The continuous function formulation allows inferring intensity values at any spatial location within the reference image.

*B. Def-Net: Spatial-temporal Learning*

We develop a spatial-temporal deformation network, Def-Net, to capture the temporal changes of the deformation fields from the patient-specific image sequence, enabling the prediction of the time-dependent deformation field at a target time point. To achieve this, the time variable $t$ is incorporated into the input of Def-Net (see Fig. 2) and the deformation network is parameterized as follows:

$$\mathcal{D}_\xi : \bar{p}_t = (p_t, t) \to u(\bar{p}_t) \text{ with } p_t \in [0,1)^3, t \in [1,T]. \quad (4)$$

In this model, the spatial-temporal coordinates $\bar{p}_t = (x, y, z, t)$ are the input, and the output of the model is the corresponding coordinate-based displacement $u(\bar{p}_t) = (\Delta x, \Delta y, \Delta z)$ at time $t$, based on which the deformation field $\phi : \mathbb{R}^3 \to \mathbb{R}^3$ mapping coordinates of target image to those of the reference image is established. That is, for each $\bar{p}_t$, $u(\bar{p}_t)$ is the displacement such that $f_t(\bar{p}_t)$ and $f_r(p_t + u(\bar{p}_t))$ correspond to the same image voxels. To avoid unphysical deformation, registration diffeomorphism is integrated in Def-Net by utilizing the stationary velocity field representation, a concept established in prior work [52]. The formulation of the deformation field is derived through the ordinary differential equation:

$$\frac{\partial \phi^{(\tau)}}{\partial t} = \upsilon(\phi^{(\tau)}) \text{ with } \phi^{(0)} = Id, \ \tau \in [0,1], \quad (5)$$

where $\phi^{(0)}$ represents the identity transformation and $\tau$ indicates time. The final deformation field $\phi^{(1)}$ is derived by integrating the stationary velocity field $\upsilon$ over the time interval $[0,1]$.

The computational flow of this network begins by encoding the spatial-temporal coordinates $\bar{p}_t$ using the Fourier feature mapping as detailed in (2), resulting $\gamma(\bar{p}_t)$. Subsequently, $\gamma(\bar{p}_t)$ is fed into Def-Net, producing the stationary velocity field $\upsilon$. Finally, the derived $\upsilon$ undergoes further processing through vector integration layers employing scaling and squaring operations [52], resulting in the generation of the ultimate deformation field $\phi$.

Our model assumes the predicted $\phi$ establishes a connection between the coordinate systems of the reference and target time points. That is, the intensity values corresponding to the coordinates after movements in the reference coordinate system should be equal to those of the initial coordinates on the target image. This connection is facilitated through the trained reference network, which provides a continuous neural representation of the reference image. To elaborate, the step involves adding the coordinate displacements $u(\bar{p}_t)$ back to the original spatial coordinates $p_t$ to form the input. Utilizing the frozen reference network ($\mathcal{R}_{\theta^*}$) trained in the last step, the resulting inference should output an image that aligns with the target image at time point $t$. This relationship constitutes the foundation for the optimization objective, guiding the training of the deformation network:

$$\xi^* = \arg\min \tfrac{1}{TN} \Sigma_{t=1}^{T} \Sigma_{i=1}^{N} \| \mathcal{R}_{\theta^*}(\mathcal{D}_\xi(\bar{p}_t^i) + p_t^i) - f_t(p_t^i) \|_2^2, \quad (6)$$

where $N$ indicates the total pixel count in the target image. Consequently, based on the given reference network, the Def-Net is trained to effectively capture the spatial-temporal neural representations through sequential image fitting.

### C. Patient-specific ROI Adaptation

Following the optimization of objective function (6), the optimized Def-Net $\mathcal{D}_{\xi^*}$ is employed to predict the deformation field at any spatial and temporal point. We apply it to monitor the contour propagation of the target anatomical regions. Assume that there is a human-annotated anatomical ROI denoted as $\mathcal{S}_r$ in the reference image $f_r$. As shown in the yellow box within Fig. 2, using the deformation field predicted by Def-Net, we can map the reference ROI $\mathcal{S}_r$ to the target ROI $\mathcal{S}_t$ at any target time point. Specifically, given a spatial-temporal point $\bar{s}_t = (s_t, t)$ on the target segmentation grid map $\mathcal{M}_t$, we can initially compute the corresponding deformed spatial coordinates $s_r = s_t + u(\bar{s}_t)$ within the reference segmentation map $\mathcal{S}_r$ using relationship (4). Note that due to the continuity of the deformation function $\phi$, the transformed coordinates in the reference segmentation map may not align precisely with integer grid locations. Thus, to determine the value at the subvoxel location $s_r$ within $\mathcal{S}_r$, nearest-neighbor interpolation is employed based on the eight neighboring voxels $\{z_r^i\}_{i=1}^{8}$ surrounding the point, namely:

$$c = \arg\min_{i \in [1,8]} \| s_r - z_r^i \|_2, \quad (7)$$

where $c$ signifies the index of the target voxel chosen for interpolation. Subsequently, the predicted segmentation value at $\bar{s}_t$ within $\tilde{\mathcal{S}}_t$ can be obtained through the following assignment:

$$\tilde{\mathcal{S}}_t(\bar{s}_t) \leftarrow \mathcal{S}_r(z_r^c). \quad (8)$$

In this way, we can effectively transfer the human-annotated patient information from the reference time point to any desired target time point through the utilization of learned spatial-temporal deformation representations.

## IV. EXPERIMENTS

To demonstrate the effectiveness of our proposed method, we conducted experiments focusing on two tasks: time-continuous image interpolation task and longitudinal deformable registration task. Notably, we validated our approach on both short-term sequential image series (e.g., 4D CT datasets in thoracic and abdominal imaging), and long-term sequential image series (e.g., longitudinal lung cancer treatment CT dataset). These diverse datasets were chosen to highlight the flexibility of our method in handling different types of temporal image data. In our study, metrics, including peak signal-to-noise ratio (PSNR), structural similarity index measure (SSIM), Dice Similarity Coefficient (Dice), and the number of voxels with a non-positive Jacobian determinant ($|J_\Phi| \leq 0$), were employed to rigorously assess the performance of the models' calculations.

### A. Datasets

#### 1) 4D CT data

We collected 4D CT data retrospectively from two patients forming Dataset A for image interpolation task. This dataset includes a pancreas 4D CT image from XXX University Hospital (Case No.1), and a lung 4D CT image from a public NSCLC dataset [53] (Case No. 2), as illustrated in Fig. 3(a). In our experiments, we selected the first phase as the reference image to learn prior embedding, and we utilized the entire series of 4D image series excluding the target image to train the deformation network. All the images underwent intensity normalization to fit within the [0,1] range. Additionally, we assessed the pairwise deformable registration performance of our adapted ST-NeRP model using the publicly available DIR-Lab dataset [54], treated as Dataset B, which comprises 4D lung CT scans for 10 patients (see Fig. 3(b)). The objective was to register the maximum inspiration image (Phase 1) to expiration image (Phase 6) for each patient, with 300 manually annotated lung landmarks available for each scan. These landmarks serve as ground-truth for evaluating target registration error (TRE) between the two extreme phases. Notably, each 4D image in both datasets consists of 3D images corresponding to 10 distinct breathing phases, captured at uniform time intervals throughout a respiratory cycle.

#### 2) Longitudinal CT data

To validate the capability of ST-NeRP in analyzing longitudinal patient data, we conducted image interpolation experiments on Dataset C, which comprises 2 patient cases, each with 8 longitudinal lung CT scans, sourced from the public NSCLC dataset [53], as depicted in Fig. 3(c). For each patient case, we designated the first 3D CT of the image sequence as the reference image for prior embedding. Furthermore, we assessed the interpolation performance on an additional Dataset D, sourced from NSCLC [53], which comprises 5 patients, each containing 5 consecutive lung CT scans with tumor delineation during lung cancer treatment (see Fig. 3(d)). We also validated the deformable registration performance of ST-NeRP on Dataset D, primarily assessed using the Dice score. All these longitudinal CT images were acquired at varying intervals over a two-month period. In the preprocessing stage, we employed affine registration tool within advanced normalization tools (ANTs) [55] to initially align the sequential CT data for all longitudinal cases respectively, establishing a consistent and reasonable foundation for the subsequent experimental validation and analysis.

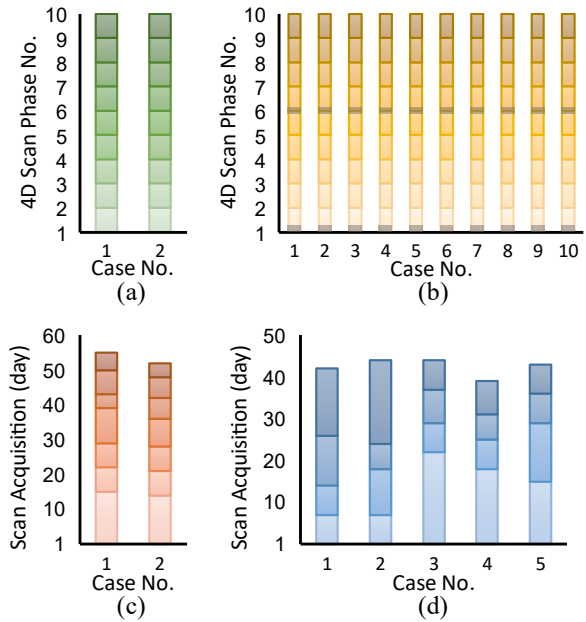

**Fig. 3.** Acquisition of medical image datasets for patient-specific study. (a) Dataset A consists of 10-phase 4D CT scans from 2 patients, focusing on pancreas and lung regions respectively in the measurement of respiratory motion. (b) Dataset B (DIR-Lab) comprises 10-phase 4D lung CT scans from 10 patients, including 300 manually annotated lung landmarks on maximum inspiration and expiration images, facilitating the evaluation of pairwise deformation registration. (c) Dataset C encompasses longitudinal images from 2 patients, each patient having 8 serial lung CT scans acquired during lung cancer treatment. (d) Dataset D contains longitudinal images from 5 patients, each patient having 5 serial lung CT scans with tumor delineation during lung cancer treatment.

*B. Implementation Details*

We implemented our networks using PyTorch [56]. In our settings, the SIRENs for both Ref-Net and Def-Net are constructed with 8 layers, each consisting of 256 neural nodes. The Fourier feature embedding for position encoding is designed with a width of 256, and the corresponding hyper-parameter $\sigma$ is set to 3. For the integration layer in Def-Net, we set the number of scaling and squaring steps to 7 following the settings elaborated in [52]. The Adam optimizer with a learning rate of 0.0001 is exploited to optimize the training objective defined in (3) for prior embedding. The training process consists of a total of 2000 iterations. Subsequently, leveraging the prior-embedded Ref-Net, Def-Net undergoes training by optimizing the objective (6) using the Adam optimizer with a learning rate of 0.00001 and the number of iterations of 2000. These hyperparameters and training settings have been carefully tuned to achieve optimal performance in our experiments.

*C. Experiments for 4D CT Interpolation on Dataset A*

In the evaluation of the ST-NeRP model's 4D CT interpolation capability using Dataset A, the first phase was used as the reference image to embed prior knowledge in Ref-Net. Def-Net was then trained with the remaining phases, excluding the target phase to be interpolated (phase No. 4, 6 & 8, respectively). As shown in TABLE I, our approach yielded exceptional prior embeddings with PSNR/SSIM of 45.26dB/0.995 and 41.30dB/0.990. Additionally, we presented the average evaluation results for both the target interpolation phase and the rest phases involved in training stage (mean$_{tgt}$/mean$_{rst}$). For Case 1 and Case 2, We observed satisfying average PSNR/SSIM values of 33.55dB/0.981 and 27.86dB/0.963, respectively, for the interpolated target phases. A notable decrease of 9% and 6% in PSNR for target phase was noted compared to the rest phase images involved in the training process. This phenomenon is further elucidated in Fig. 4(a), which illustrates the evaluation on the entire phase series using the well-trained ST-NeRP. Such a decline can be attributed to the absence of visibility of the target phase during the training stage and the intricate nonlinearity in the medical image series. Nevertheless, we observed a slight and acceptable decrease of 0.7% and 1.3% in SSIM, indicating that despite being influenced by intensity noise, the interpolation results still adequately predict structural characteristics. Notably, the predicted deformation fields exhibit diffeomorphic properties across all the cases, demonstrating pronounced topology preservation, invertibility and biomechanical plausibility. Qualitative examples were provided to visualize the performance in 4D CT interpolation tasks, as shown in Fig. 5, where the columns of the specific patient case show the different cross-sectional slices of the corresponding 3-D volumes. The first and second rows display the ground truth (GT) reference image and the generated reference image, respectively, at different slices. The visual comparison demonstrates that the generated reference image closely aligns with the ground truth, indicating the successful acquisition of a highly accurate prior embedding through Ref-Net training. The third and fourth rows depict the GT target image and the predicted target image, respectively. The corresponding diffeomorphic deformation field, intuitively illustrating the "movement" between reference and target phases, is displayed in the last row. Our findings reflect that our model possesses an exceptional capability to accurately represent the reference image through Ref-Net and showcases promising performance in image interpolation, as depicted in Fig. 5. We will further conduct quantitative analysis to assess our model's effectiveness in capturing structural variations in the patient's tumor regions in Section IV.D.

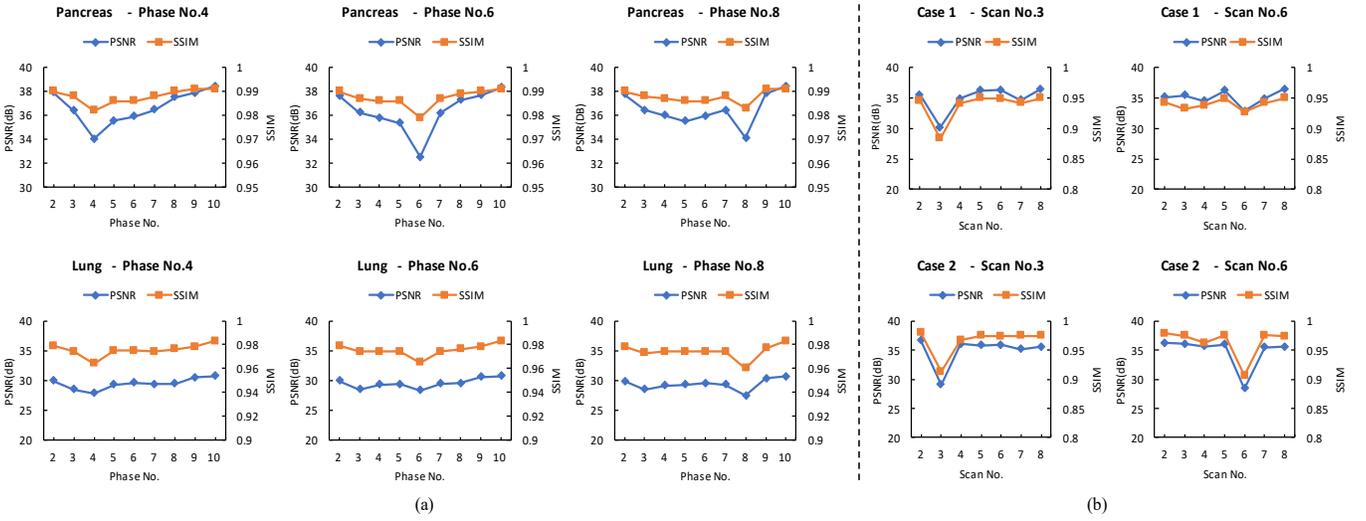

**Fig. 4.** Evaluation for medical image interpolation using PSNR and SSIM metrics in patient-specific serial imaging study. (a) Evaluation results for CT image interpolation at target phases of pancreas and lung 4D CT scans, respectively, in Dataset A. (b) Evaluation results for lung CT image interpolation at target time points in the longitudinal imaging study within Dataset C.

TABLE I
QUANTITATIVE RESULTS FOR 4D CT INTERPOLATION ON DATASET A ($MAEN_{TGT}$ / $MEAN_{RST}$)

| Case/Scan No. | | PSNR (dB) ↑ | SSIM ↑ |
|---|---|---|---|
| 1 | 1(prior) | 45.26 | 0.995 |
| | 4 | 34.05/37.03 | 0.982/0.989 |
| | 6 | 32.49/36.83 | 0.979/0.988 |
| | 8 | 34.10/36.81 | 0.983/0.988 |
| | **Avg.** | **33.55/36.89 (↓9%)** | **0.981/0.988 (↓0.7%)** |
| 2 | 1(prior) | 41.30 | 0.990 |
| | 4 | 27.86/29.67 | 0.964/0.977 |
| | 6 | 28.33/29.67 | 0.965/0.977 |
| | 8 | 27.39/29.57 | 0.960/0.975 |
| | **Avg.** | **27.86/29.64 (↓6%)** | **0.963/0.976 (↓1.3%)** |

TABLE II
QUANTITATIVE RESULTS FOR LONGITUDINAL CT INTERPOLATION ON DATASET B ($MAEN_{TGT}$ / $MEAN_{RST}$)

| Case/Scan No. | | PSNR (dB) ↑ | SSIM ↑ |
|---|---|---|---|
| 1 | 1(prior) | 45.65 | 0.994 |
| | 3 | 30.17/35.72 | 0.884/0.946 |
| | 6 | 32.84/35.46 | 0.927/0.943 |
| 2 | 1(prior) | 44.41 | 0.991 |
| | 3 | 29.02/35.82 | 0.913/0.974 |
| | 6 | 28.41/35.78 | 0.906/0.973 |
| **Avg.** | | **30.11/35.70 (↓15.7%)** | **0.908/0.959 (↓5.3%)** |

*D. Experiments for Longitudinal CT Interpolation*

*1) Dataset C: 2 patients each with 8 serial CT scans*

We conducted further investigations into the performance of ST-NeRP on Dataset C, which includes two longitudinal lung CT cases, each comprising 8 sequential 3D CT scans. The initial CT scan was designated as the reference, represented effectively by our trained Ref-Net as a continuous function parametrized by network weights. Subsequently, we proceed to learn a spatial-temporal continuous Def-Net to predict the target image at any other time point. Here, we aimed to predict scan No. 3 and No. 6 respectively, positioned at intermediate stage within the scan sequence, utilizing the remaining CT scan images within the series for model training. The obtained PSNR and SSIM values for the predicted target images are presented in TABLE II, indicating an average PSNR of 30.11dB and SSIM of 0.908. Additionally, an evaluation of the model's performance across the entire image series was conducted to identify variations among different temporal points, as shown in Fig. 4(b), similar to the outcomes discussed in Section IV-C.

Fig. 6 presents the outcomes of the interpolation process at scan No. 6 within the longitudinal lung CT series. The columns in row 4 represent the different cross-sectional slices of the target 3D volume, illustrate the proficient generation of high-quality images via our interpolation methodology, demonstrating the effectiveness of our approach in delivering reliable outcomes in the realm of interpolation tasks. Likewise, analogous diffeomorphic deformation fields were obtained. However, it is noteworthy to observe that the cross-sectional slices related to the deformation fields exhibit a higher degree of intricacy, especially beyond the thoracic cavity region, compared to the intricacy observed in the context of the 4D CT interpolation task. This discernible discrepancy could potentially stem from inherent challenges, e.g., initial alignment process across distinct scanning instances.

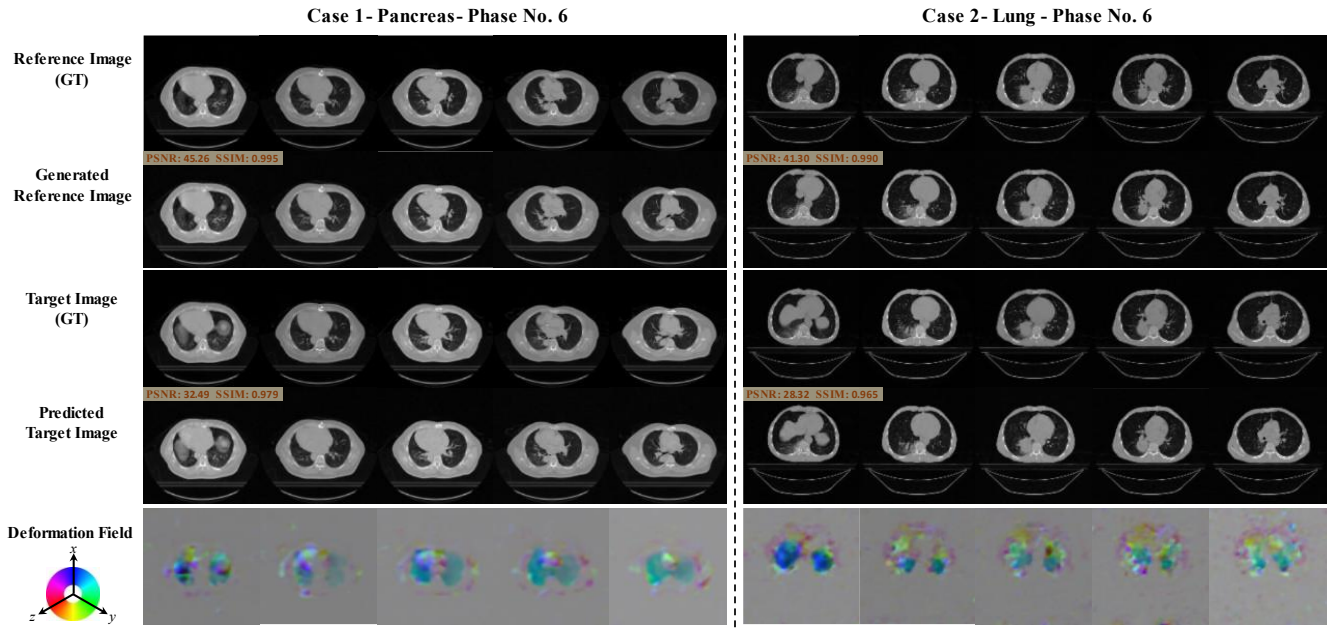

**Fig. 5.** Qualitative examples for medical image interpolation at the target time points (e.g., Phase No. 6) in 4-D CT imaging on Dataset A, one for pancreas region shown on the left and the other for lung region shown on the right. The generated reference images predicted by Ref-Net with prior embedding learned from ground truth (row 1) are shown in row 2. The target image ground truth is shown in row 3, followed by the predicted target images in row 4. Row 5 showcases the corresponding deformation fields, where the spatial dimension x, y and z are mapped to each of the BGR color channels, respectively. The columns represent different cross-sectional images of the corresponding entire 3-D volumes.

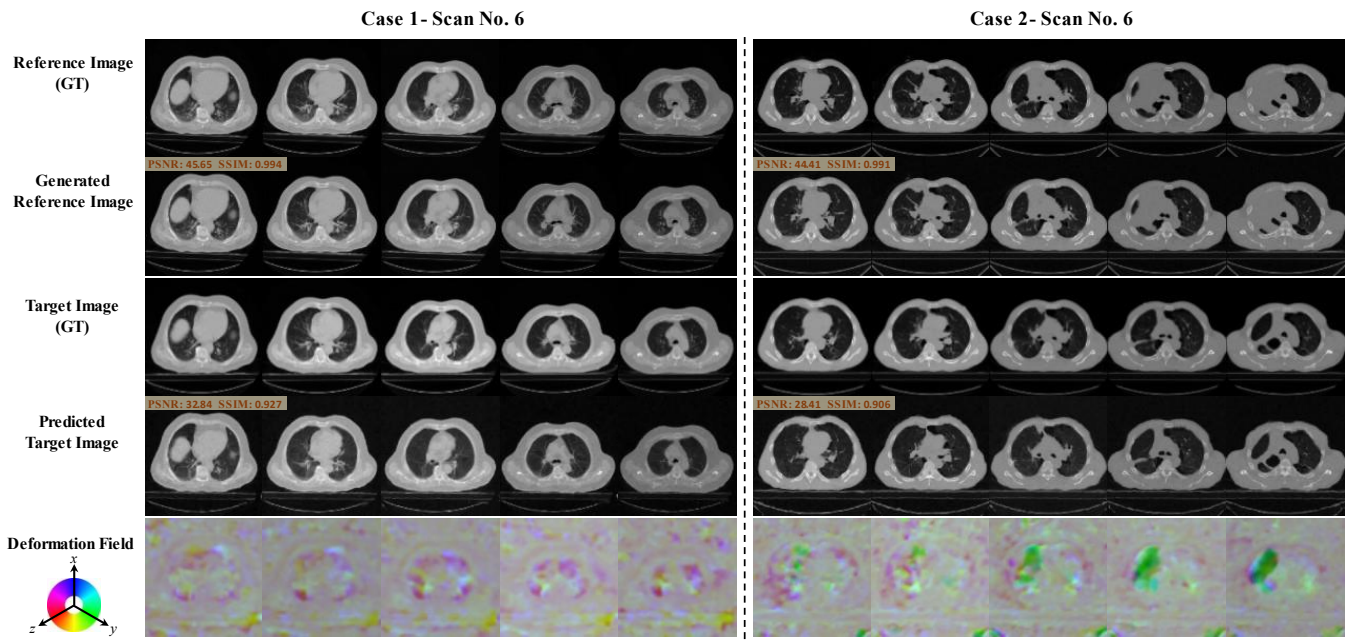

**Fig. 6.** Qualitative examples for medical image interpolation at the target time points (e.g., Scan No. 6) in patient-specific lung-region longitudinal imaging study on Dataset C. The generated reference images predicted by Ref-Net with prior embedding learned from ground truth (row 1) are shown in row 2. The target image ground truth is shown in row 3, followed by the predicted target images in row 4. Row 5 showcases the corresponding deformation fields, where the spatial dimension x, y and z are mapped to each of the BGR color channels, respectively. The columns represent different cross-sectional images of the corresponding entire 3-D volumes.

TABLE III
QUANTITATIVE RESULTS FOR LONGITUDINAL CT INTERPOLATION ON DATASET D (MEAN)

| Tasks | Method | Fraction 3 | | | | Fraction 4 | | | |
|---|---|---|---|---|---|---|---|---|---|
| | | PSNR (dB)↑ | SSIM↑ | Dice↑ | $|J_\Phi| \leq 0$↓ | PSNR (dB)↑ | SSIM↑ | Dice↑ | $|J_\Phi| \leq 0$↓ |
| Time-dependent Registration | ST-NeRP | 37.52 | 0.984 | 0.778 | **0** | 37.61 | 0.985 | 0.787 | **0** |
| | ST-NeRP (w/o prior) | **39.83** | **0.990** | 0.774 | 0 | **39.98** | **0.991** | 0.787 | 0 |
| | ST-NeRP (w/o t) | 38.12 | 0.986 | **0.786** | 0 | 38.14 | 0.976 | **0.789** | 0 |
| Time-continuous Interpolation | ST-NeRP | 30.06 | 0.937 | **0.700** | **0** | 29.63 | 0.931 | **0.694** | **0** |
| | ST-NeRP (w/o prior) | **30.57** | **0.944** | 0.673 | 0 | **29.92** | **0.942** | 0.684 | 0 |

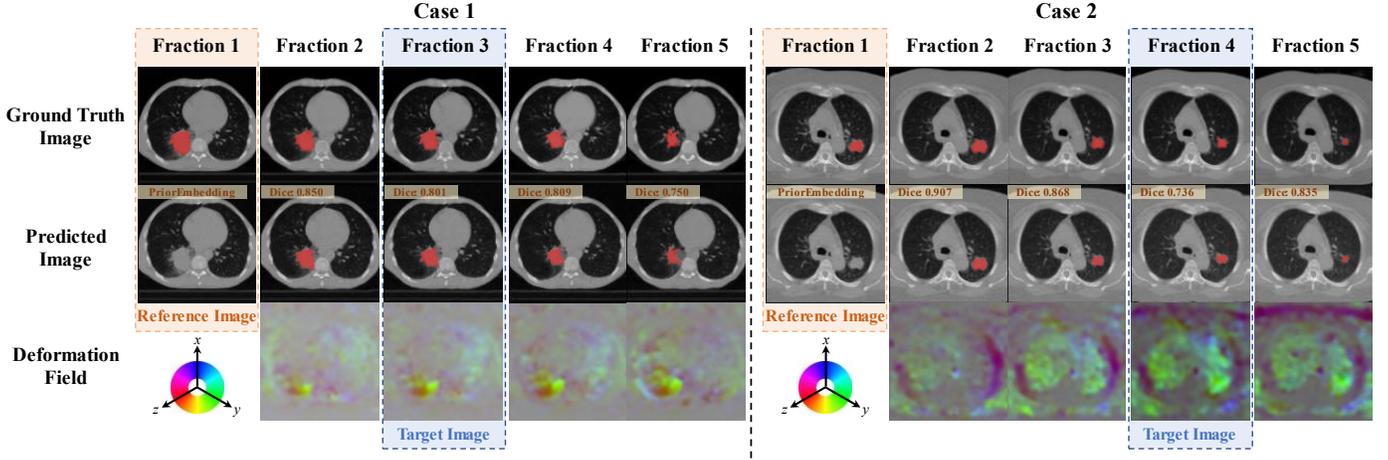

**Fig. 7.** Experimental examples for longitudinal image interpolation along with tumor region estimation (Fraction 3 for Case 1 and Fraction 4 for Case 2) on Dataset D, consisting of 5 patients undergoing lung cancer treatment. The time-dependent deformable registration results on those longitudinal sequence excluding the target interpolation image are also presented. Tumor regions are indicated by red masks and the corresponding deformation fields are shown alongside for reference, where the spatial dimension x, y and z are mapped to each of the BGR color channels, respectively.

*2) Dataset D: 5 patients each with 5 serial CT scans*

To delve deeper into the interpolation performance of our method for monitoring the anatomical changes in patient-specific study, we conducted experiments on Dataset D, which comprises 5 patients, each containing 5 serial lung CT scans with tumor delineations obtained during lung cancer treatment. Similarly, we selected the initial CT scan as the reference and performed interpolations at fractions 3 and 4, respectively, for each patient. Notably, we not only present the time-continuous interpolation results at specific target fractions but also demonstrate the time-dependent deformation registration performance across all the other fractions used for training. To showcase the effectiveness of our framework, we conducted a comprehensive ablation study to demonstrate the superiority of our ST-NeRP method, as shown in TABLE III. We compared our approach with ST-NeRP (w/o prior), which utilizes STL for deformable transformation instead of Ref-Net, essentially employing a linear interpolation operation. Our findings indicate that ST-NeRP outperforms in terms of Dice score, while exhibiting lower performance in PSNR and SSIM. This discrepancy can be attributed to differences in intensity-based transformation and optimization strategy. In ST-NeRP (w/o prior), the integration of linear interpolation within STL leads to smoother output images, resulting in higher PSNR and SSIM scores. However, in this scenario, the intensity-based loss function may not effectively guide the network to accurately identify corresponding points between the output and ground truth images, especially when intensities around the predicted point are very similar after linear smoothing operation. In contrast, the Ref-Net within ST-NeRP, utilizing implicit neural representation to continuously represent the reference image, can help to predict the deformed points along with their intensity values directly, without relying on linear interpolation to smooth intensity. Consequently, it ensures better structural accuracy, slightly higher in the time-dependent registration, and significantly improved in our interpolation task (Dice score: Fraction 3 - 0.700 vs. 0.673; Fraction 4 - 0.694 vs. 0.684). Fig. 7 presents selected examples showcasing our method's image interpolation at Fraction 3 and Fraction 4, respectively. Additionally, we compared our method with ST-NeRP (w/o t), which focuses on pairwise registration by simplifying the input of Def-Net into a specific image instead of the image series. We observed highly comparable results to ST-NeRP (w/o t) in terms of all metrics, with just a slight decrease. This consistency underscores the capability of ST-NeRP to adeptly learn both temporal and spatial representations.

*E. Experiments for Deformable Registration*

*1) Dataset B: DIR-Lab*

10To elucidate the versatile capabilities of ST-NeRP, we customized the model for deformable registration tasks. This adaptation simplified the input configuration of Def-Net, focusing solely on the target image rather than multiple consecutive images. We conducted comparisons with various existing methods, encompassing both learning-based and traditional optimization-based approaches, using the 4D CT dataset DIR-LAB. This evaluation involved computing the TRE for 300 predefined anatomical landmark pairs on two extreme phase images (maximum inspiration-to-expiration registration) for each patient, as outlined in TABLE IV. Specifically, we compared our method with two state-of-the-art INR-based registration techniques: IDIR [22] and an enhanced version incorporating cycle-consistency proposed by Van Harten et al. [48]. Additionally, we evaluated our method against two CNN-based methods, namely DLIR [57] and an anatomically constrained CNN proposed by Hering et al. [43]. Furthermore, we included comparisons with two traditional optimization-based deformable registration approaches: the deformable registration tool integrated within ANTs [55] and Uniform B-Splines [58]. The results for the two INR-based methods were sourced from [48], while the other methods' results were taken from their respective papers, except for ANTs, for which we generated the result using its toolbox. Our findings reveal that our ST-NeRP method attained the average registration error of $1.06 \pm 1.23$ mm, outperforming all the traditional optimization-based and CNN-based methods. Additionally, our method surpasses IDIR (1.10±1.42 mm) and achieves comparable performance to the cycle-consistent INR proposed by Van Harten et al (1.04±1.11 mm), demonstrating the excellent registration performance of our method.

*2) Dataset D: 5 patients each with 5 serial CT scans*

To further underscore the efficacy of our ST-NeRP, we conducted validation on Dataset D for longitudinal registration analysis, primarily assessing it through the Dice score metric. In accordance with clinical requirements, our experimental efforts encompassed two distinct schemes across 5 patients: long-range deformable registration, aiming to predict the target CT scan $f_t$ given the initial CT scan $f_1$ as reference time point, and short-range deformable registration, targeting prediction between two adjacent CT scans $f_t$ and $f_{t-1}$, using the last neighboring scan as a reference. We compared our approach with the INR-based deformable registration approach IDIR [22] the widely-used open-source deformable registration tool integrated within ANTs [55], as shown in Fig. 8. In the context of the long-range registration ( $f_1 \rightarrow f_t$ ), a discernible decrease in the Dice score was observed. This diminishing trend can be attributed to the evolving nature of the treatment process, where the tumor undergoes substantial anatomical modifications, introducing complexities for deformable registration, particularly when referencing back to the initial time point. In contrast, the short-range registration ( $f_{t-1} \rightarrow f_t$ ) showed comparatively improved outcomes due to the gradual progression of image deformations between successive scans, resulting in better results overall. Moreover, we conducted a thorough quantitative analysis employing diverse assessment metrics, as outlined in TABLE V. It collectively demonstrates a comprehensive improvement compared to ANTs across both schemes, notably pronounced in the long-range scenario with a significant enhancement of 2.1% in the Dice score. Specifically, the Average Dice scores for ST-NeRP and ANTs are 0.791 and 0.770 in the context of long-range registration, and 0.803 and 0.789 in short-range registration. Furthermore, our method exhibited comparable PSNR and SSIM results compared to IDIR, and demonstrated superior Dice scores, particularly evident in long-range registration with a 1.2% improvement, showcasing its strong capability in estimating anatomical structures. Moreover, we conducted an ablation study by comparing with ST-NeRP (w/o prior), which showed the highest values in PSNR and SSIM, attributed primarily to its linear interpolation operation of STL, resulting in smoother output images. Nevertheless, ST-NeRP still outperformed in Dice score due to the powerful prior embedding of Ref-Net. Selected examples for visualization are illustrated in Fig. 9 for long-range registration and Fig. 10 for short-range registration. The first two columns showcase the GT images and corresponding labels, while the third, fifth and seventh columns present predicted target images and labels for ST-NeRP, IDIR and ANTs, respectively. The corresponding deformation fields are displayed alongside for reference. Notably, ST-NeRP demonstrated a significant advancement in the structural quality of warped images. For instance, in Example 1 within Fig. 10, our approach provided a more accurate estimation of tumor contour at the target time point, evidenced by a Dice score of 0.874. In contrast, IDIR and ANTs struggled to capture the deformation, resulting in lower Dice scores of 0.846 and 0.850, respectively. An additional virtue of our model was its achievement of diffeomorphic registration, evidenced through the generation of smooth deformation fields devoid of folding voxels ($|J_\Phi| \leq 0$), providing a significant advantage over ANTs. In general, the validation of tumor contour estimation showcased the efficacy of the learned deformation field via our algorithm, underlining the superiority of ST-NeRP in the domain of patient-specific deformable registration tasks and highlighting its potential to facilitate patient-adaptive radiation therapy.

TABLE IV
QUANTITATIVE RESULTS FOR DEFORMABLE REGISTRATION ON DATASET B (DIR-LAB)

| Method | mean ± std (mm) |
|---|---|
| ANTs [55] | 3.25±3.56 |
| Uniform B-Splines [58] | 1.36±0.99 |
| DLIR [57] | 2.64±4.32 |
| CNN with anatomical constraints [43] | 1.14±0.76 |
| IDIR [22] | 1.10±1.42 |
| Cycle-consistent INR [48] | 1.04±1.11 |
| ST-NeRP | 1.06±1.23 |

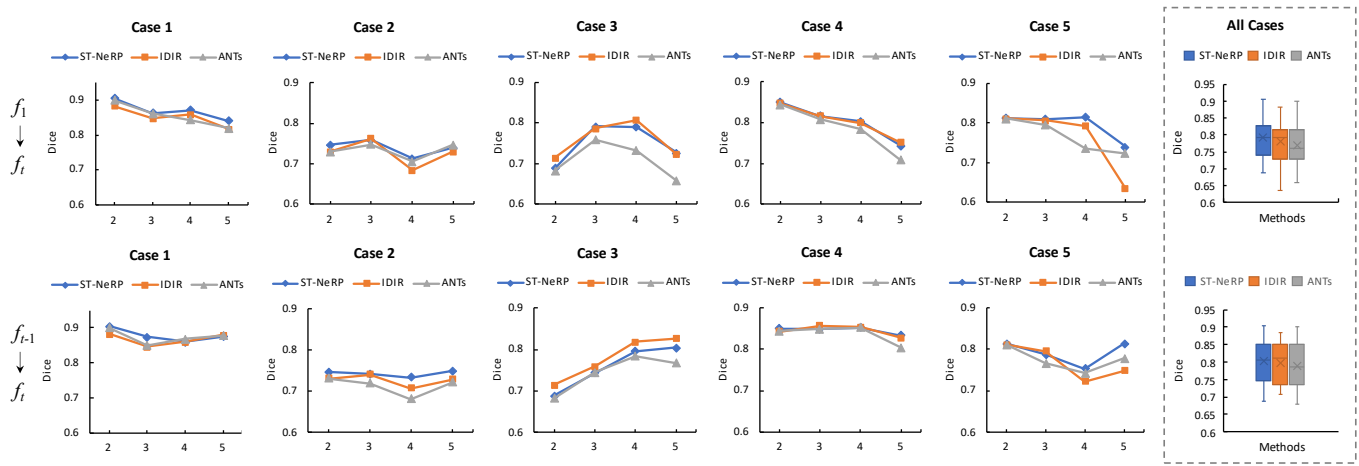

**Fig. 8.** Quantitative results of ST-NeRP, IDIR and ANTs evaluated with Dice score for long-range ($f_1 \to f_t$) and short-range ($f_{t-1} \to f_t$) patient-specific lung-region deformable registrations on Dataset D which includes 5 patient cases.

TABLE V
QUANTITATIVE RESULTS FOR DEFORMABLE REGISTRATION ON DATASET D (MEAN)

| Scheme | Method | PSNR (dB)↑ | SSIM↑ | Dice↑ | $|J_\Phi| \leq 0$↓ |
|---|---|---|---|---|---|
| $f_1 \to f_t$ | ANTs [55] | 33.04 | 0.969 | 0.770 | 5699 |
| | IDIR [22] | 38.73 | 0.988 | 0.780 | 0 |
| | ST-NeRP | 38.21 | 0.986 | **0.791** | **0** |
| | ST-NeRP (w/o prior) | **40.60** | **0.992** | 0.783 | 0 |
| $f_{t-1} \to f_t$ | ANTs [55] | 33.10 | 0.975 | 0.789 | 6911 |
| | IDIR [22] | 38.50 | 0.988 | 0.798 | 0 |
| | ST-NeRP | 38.83 | 0.988 | **0.803** | **0** |
| | ST-NeRP (w/o prior) | **40.10** | **0.991** | 0.802 | 0 |

## V. DISCUSSION

A comprehensive understanding of the disease progression through the analysis of monitored longitudinal data, including CT and MRI scans, is pivotal in clinical practice for informed decision-making and improved patient care. This paper introduces ST-NeRP, a personalized learning framework to explore the continuous deformation patterns inherent in sequential medical data, where the spatial-temporal features are effectively extracted by the implicit neural representation learning. Our approach offers two significant advantages, including: 1) patient-specific attribute without the need for extensive data for model training; and 2) applicability to both image interpolation and deformable registration tasks. To demonstrate the efficacy of ST-NeRP, we conducted validation experiments on 4D CT images and longitudinal CT series focusing on thoracic and abdominal regions, and it has great potential to be generalizable across diverse imaging modalities commonly encountered in the clinical settings. To better understand the role of Ref-Net within our framework, we conducted an ablation study analyzing prior embeddings (see TABLE III and IV). The comprehensive comparison, as illustrated in TABLE VI, showcases the impact of Ref-Net on both interpolation and registration tasks in the longitudinal study involving five patients. We observe that our method improves Dice scores in both tasks, especially in image interpolation by 1.8% = 0.697-0.679, demonstrating the effective feature representation of Ref-Net and its substantial contribution to the overall model performance. Although ST-NeRP (w/o prior) exhibits slightly superior performance in PSNR and SSIM, attributed to the smoothness property of integrated STL module, we prioritize ST-NeRP's advantage in better capturing anatomical structure information, which usually holds greater significance in clinical decision-making, tumor contour propagation and treatment plan adaptation.

Our method excels in personalized analysis for precision medicine, circumventing common issues associated with data-driven learning methods, such as the need for extensive training samples and the generalization gap between training and test data. However, it requires optimizing our model from scratch for each patient, which can be inefficient. Therefore, it is crucial to explore promising solutions to address this challenge. This includes capturing shared features across patients' data or exploring transfer learning techniques to adapt pre-trained models from one patient to new patients with minimal retraining. Another alternative approach involves combining the advantages of both data-driven and INR methods. This entails utilizing a pre-trained data-driven network to predict an initial deformation quickly, thus simplifying the search space for ST-NeRP to estimate an optimal residual deformation. This hybrid strategy has the potential to enhance the model's accuracy, efficiency, and generalizability [49]. Additionally, we recognize the limitations of the exclusive reliance on image intensity for model learning, which can lead to inaccuracy in deformation estimation within intensity-homogeneous regions, e.g., closely contiguous tumor and heart in thoracic CT imaging. Consequently, our future endeavors will be directed towards exploring the integration of supplementary biomarkers, such as ROI contours or anatomical landmarks, as optimization constraints within the training process.

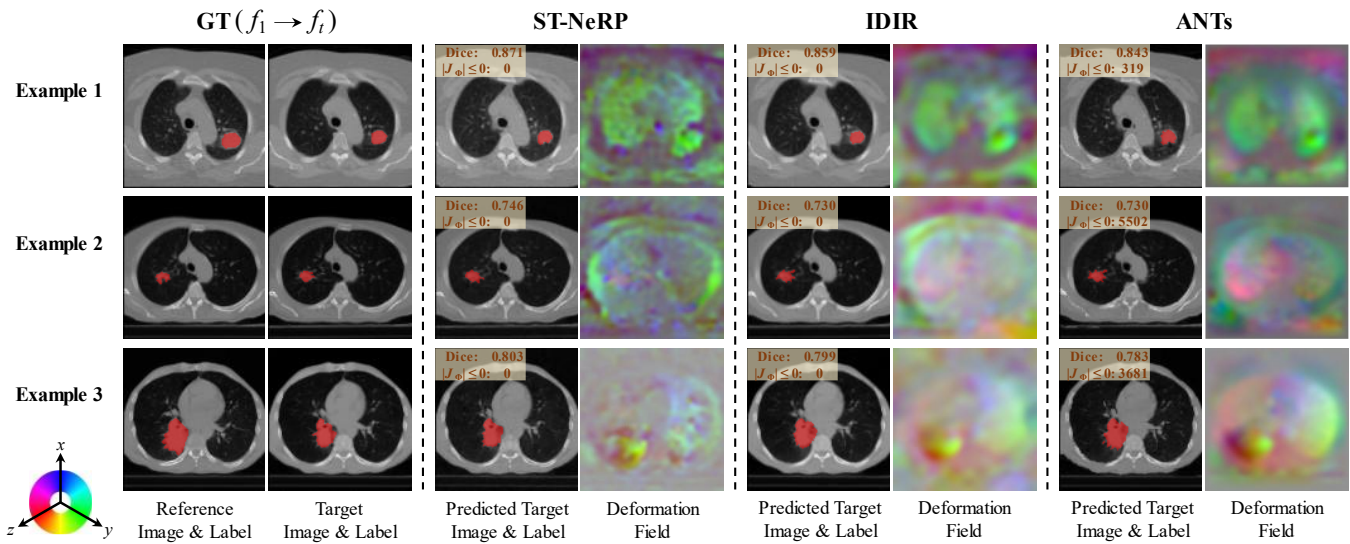

**Fig. 9.** Comparison examples of ST-NeRP, IDIR and ANTs for long-range patient-specific lung-region registration on Dataset D. The predicted target images and segmentation labels are presented for qualitative evaluation, along with the corresponding deformation field, where the spatial dimension x, y and z are mapped to each of the BGR color channels, respectively. The tumor regions are delineated with red masks for better visualization. The quantitative results are also provided for reference including Dice score and the number of voxels with a non-positive Jacobian determinant.

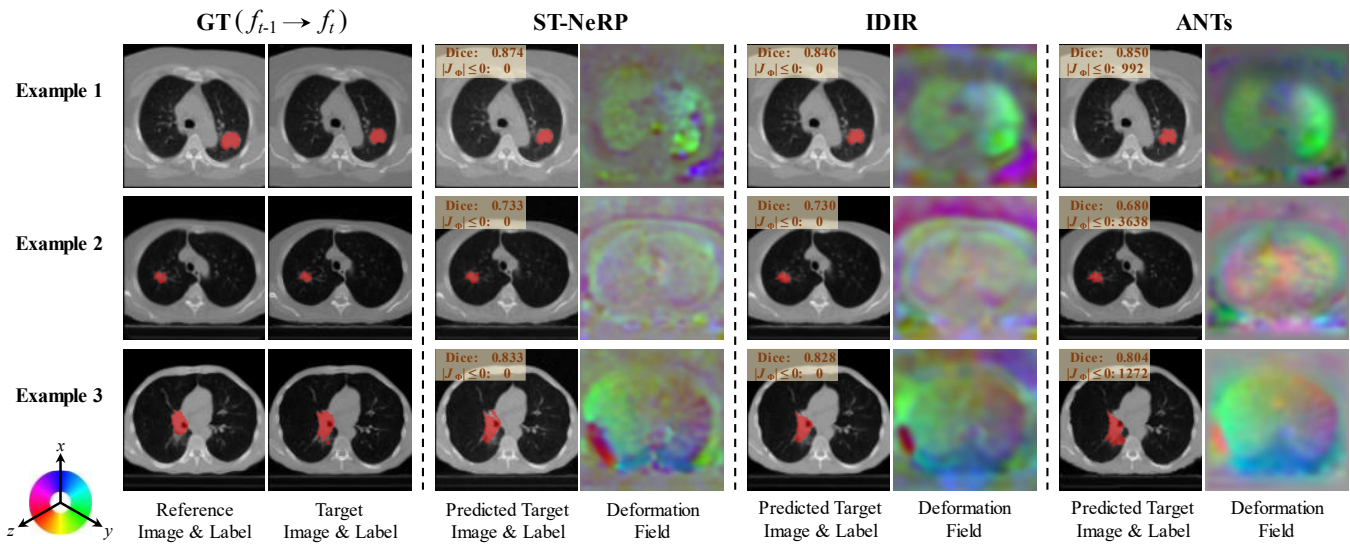

**Fig. 10.** Comparison examples of ST-NeRP, IDIR and ANTs for short-range patient-specific lung-region registration on Dataset D. The predicted target images and segmentation labels are presented for qualitative evaluation, along with the corresponding deformation field, where the spatial dimension x, y and z are mapped to each of the BGR color channels, respectively. The tumor regions are delineated with red masks for better visualization. The quantitative results are also provided for reference including Dice score and the number of voxels with a non-positive Jacobian determinant.

## VI. CONCLUSION

TABLE VI
PRIOR EMBEDDING ANALYSIS IN LONGITUDINAL CT INTERPOLATION ON DATASET D (MEAN)

| Tasks | Time-continuous Interpolation | | | | Time-dependent Registration | | | |
|---|---|---|---|---|---|---|---|---|
| | PSNR (dB)↑ | SSIM↑ | Dice↑ | $|J_\Phi| \leq 0$↓ | PSNR (dB)↑ | SSIM↑ | Dice↑ | $|J_\Phi| \leq 0$↓ |
| ST-NeRP | 29.85 | 0.934 | **0.697** | **0** | 38.52 | 0.987 | **0.797** | **0** |
| ST-NeRP (w/o prior) | **30.25** | **0.943** | 0.679 | 0 | **40.35** | **0.992** | 0.793 | 0 |

In this study, we proposed a novel ST-NeRP model for patient-specific 4D or longitudinal imaging study. Our method effectively learns a spatial-temporal deformation function by integrating prior knowledge of the reference image, which proves instrumental in monitoring subtle alterations in target images. Through comprehensive experiments using both 4D CT and longitudinal image datasets in thoracic and abdominal regions, we successfully demonstrated the capability of the ST-NeRP approach in performing time-continuous image interpolation across discrete series of images. Additionally, our model exhibited excellent performance in deformable registration and reliable adaptation inferences regarding anatomical ROI regions, i.e., contour propagation. Given the ubiquity of multiple intermittent imaging scans in clinical practice, the proposed technique should find widespread applications in the future.